\documentclass[twoside]{article}
\usepackage{fleqn,espcrc2}

\usepackage{graphicx}

\usepackage{latexsym}

\newcommand{\AmS}{{\protect\the\textfont2
  A\kern-.1667em\lower.5ex\hbox{M}\kern-.125emS}}

\title{Nucleon axial charge from quenched lattice QCD
with domain wall fermions and improved gauge action
\thanks{Presented by S.\ Sasaki}}

\author{
Shoichi Sasaki\address[TOKYO]{Department of
Physics, University of Tokyo, Hongo 7-3-1, Bunkyo-ku, Tokyo 113-0033, 
Japan.},
Tom Blum\address[RBRC]{RIKEN-BNL Research Center,
Brookhaven National Laboratory, Upton, NY 11973-5000, USA.},
Shigemi Ohta\address[KEK]{Institute for Particle and Nuclear Studies, KEK,
Tsukuba, Ibaraki 305-0801, Japan.}\addressmark[RBRC]
and Kostas Orginos\addressmark[RBRC] 
[RBC Collaboration]
}
       
\def\Ga    {g_{_{A}}}
\def\Gv    {g_{_{V}}}

\def\Za    {Z_{_A}}
\def\Zv    {Z_{_V}}
\def\beqn{\begin{equation}}
\def\eeqn{\end{equation}}
\def\barr{\begin{eqnarray}}
\def\earr{\end{eqnarray}}
\begin{document}

\begin{abstract}
In our previous DWF calculation with the Wilson gauge action at $\beta=6.0$
($a^{-1}\simeq$ 1.9 GeV) on a $16^3 \times 32 \times 16$ lattice, we
found that $\Ga$ had a fairly strong dependence on the quark mass.
A simple linear extrapolation of $\Ga$ to the chiral limit yielded a 
value that was almost a factor of two smaller than the experimental one.  
Here we report our recent study of this issue. 
In particular, we investigate possible errors arising from 
finite lattice volume, especially in the lighter quark mass region. 
We employ a RG-improved gauge action (DBW2), which maintains 
very good chiral behavior even on a coarse lattice 
($a^{-1}\simeq$ 1.3 GeV), in order to perform simulations at large
physical volume ($> (2{\rm fm})^3$).  Our preliminary results suggest that
the finite volume effect is significant.
\vspace{1pc}
\end{abstract}

\maketitle

The nucleon (iso-vector) axial charge $\Ga$ is a particularly interesting
quantity.  We know precisely the experimental value $\Ga=1.2670(35)$
from neutron beta decay.  
Deviation of this quantity from unity, in contrast to the vector charge,
$\Gv=1$, reflects the fact that the axial current is only partially 
conserved in the strong interaction while the vector current is 
exactly conserved. However neither lattice-QCD nor any model calculation 
have successfully reproduced this value.
Thus, the calculation of $\Ga$ is an especially relevant test of
the chiral properties of DWF in the baryon sector.
In addition, calculation of $\Ga$ is an important first step 
in studying polarized nucleon structure functions since
$\Ga=\Delta u - \Delta d$ where 
$\langle p,s|{\bar q_{f}}\gamma_{5}\gamma_{\mu}q_{f}|p,s\rangle
=2s_{\mu}\Delta q_{f}$ with $s^2=-1$ and $s\cdot p=0$.

We follow the standard practice \cite{all} for the
calculation of $\Gv$ and $\Ga$. 
We define the three-point functions for the relevant components of the 
local vector current  $J_{_{V}}^{f}={\bar q}_{f}\gamma_{4}q_{f}$ 
and axial current
$J_{_{A}}^{f}={\bar q}_{f}\gamma_{5}\gamma_{i}q_{f}$ ($i=1,2,3$):
%
%
\beqn
G_{_\Gamma}^{u, d}(t,t')={\rm Tr}[P_{_\Gamma}\sum_{{\vec x},{\vec x}'}
\langle T N(x)J_{_{\Gamma}}^{u,d}(x'){\bar N}(0)\rangle]
\\
\eeqn
where $\Gamma$ = $V$ (vector) or $A$ (axial) with
$P_{_{V}}$ = $(1+\gamma_{4})/2$ and 
$P_{_{A}}$ = $P_{_V}\gamma_{i}\gamma_{5}$.
We use the nucleon interpolating operator
$N=\varepsilon_{abc}(u_{a}^{T}C\gamma_{5}d_{b})u_{c}$.
For the axial current, the three-point function is averaged over 
$i=1,2,3$.  The lattice estimates of vector and axial charges can
be derived from the ratio between two- and three-point functions
%
%
\beqn
g_{_{\Gamma}}^{\rm 
lattice}={{G_{\Gamma}^{u}(t,t')-G_{\Gamma}^{d}(t,t')}
\over G_{_{N}}(t)}\;,
\eeqn
where
$G_{_{N}}(t)={\rm Tr}[P_{+}\sum_{\vec x}\langle T N(x){\bar
N}(0)\rangle]$.  Recall that in general lattice operators ${\cal
O}_{\rm lat}$ and continuum operator
${\cal O}_{\rm con}$ are regularized in different schemes.  The 
operators are related by a renormalization factor $Z_{\cal O}$:
${\cal O}_{\rm con}(\mu)=Z_{\cal O}(a\mu){\cal O}_{\rm lat}(a)$. 
This implies that the continuum value of vector and axial charges
are given by $g_{_{\Gamma}}=Z_{_{\Gamma}}g_{_{\Gamma}}^{\rm 
lattice}$.  In the case of conventional Wilson fermions, the 
renormalization factor $\Za$ is usually estimated in  
perturbation theory ($\Za$ differs from unity because 
of explicit symmetry breaking).  For DWF, the conserved axial current 
receives no renormalization.  This is not true for the lattice local 
current.  An important advantage with DWF, however, is that the 
lattice renormalizations, $\Zv$ and $\Za$, 
of the local currents are the same \cite{rbc01} 
so that the ratio $(\Ga/\Gv)^{\rm lattice}$ directly yields 
the continuum value $\Ga$ \cite{sbo00}.

Our first DWF results are analyzed on 200 quenched gauge 
configurations at $\beta=6.0$ on a $16^3 \times 32 \times 16$ 
lattice with $M_{5}=1.8$ \cite{sbo00}.  We found
$\Ga/\Gv$ exhibited a strong dependence on the 
quark mass \cite{sbo00}.  
A simple linear extrapolation of $\Ga$ to the 
chiral limit yielded a value that was a factor of two smaller 
than the experimental one.  This issue requires
checking related systematic effects arising from finite lattice
volume and quenching (for example quenched chiral logarithms,
zero modes, and the absence of the full pion cloud). 
In this work, we mainly focus on the former.
Indeed, the above calculation was employed in a rather small
physical volume $\sim(1.6 {\rm fm})^3$ in comparison with
the proton charge radius $\sim 0.7 {\rm fm}$.

To determine $\Ga$ in large physical volume $> (2 {\rm fm})^3$,
we perform our simulations on a coarser lattice.
However, it is difficult to maintain good chiral properties
on a coarse lattice at fixed $L_{s}$ with the Wilson gauge
action.  Recent studies have shown that the Iwasaki gauge
action enables studies of quenched DWF at smaller $L_{s}$ 
than the Wilson gauge action \cite{rbc00,cp-pacs}. 
In this work, we employ a similar
type of renormalization group improved gauge action,
DBW2 ($c_{1}=-1.4069$) \cite{qcdtaro}:
%
%
\begin{equation}
S_{G}\propto c_{0}\sum_{\rm plaq} {\rm Tr}P(1\times1) + c_{1}\sum_{\rm 
rect} {\rm Tr}P(1\times 2)
\end{equation}
with $c_{0}+8c_{1}=1$.  The chiral symmetry
of DWF with DBW2 is significantly improved over 
the Iwasaki action \cite{kostas} and also
provides good scaling behavior of the light hadron 
spectrum \cite{yaoki}.

Numerical simulations are performed 
at $\beta =0.87$ ($a\approx 0.15$ fm)
on lattice sizes $8^3 \times 24 \times 16$ and 
$16^3 \times 32 \times 16$ with $M_{5}=1.8$.
Our preliminary results are analyzed on 170 quenched 
gauge configurations for the smaller lattice 
and 86 configurations for the larger lattice.

First, we check whether or not $\Zv=\Za$ is well satisfied
on this coarse lattice. 
The vector renormalization derived from a plateau of 
$1/\Gv^{\rm lattice}$ plotted against the location of current 
insertions.
%
%
\begin{figure}[t]
\includegraphics[width=70mm]{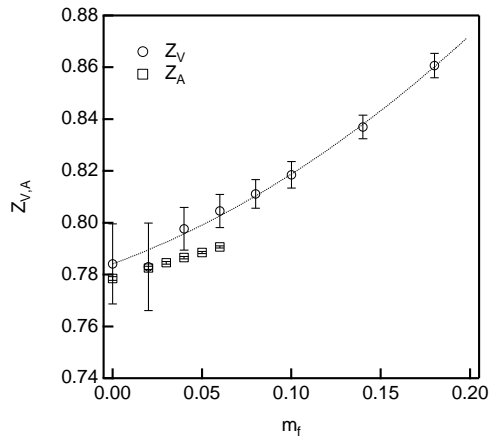}
\vskip -.2in
\caption{Quark mass dependence of the 
vector renormalization, \(Z_{_V} = 1/g_{_V}^{\rm
lattice}\).  A quadratic extrapolation
yields the value of 0.784(15) at \(m_f\)=0, which
quite agrees with the value $\Za=0.7777(4)$~\cite{yaoki}.}
\label{fig:zv_ave}
\vskip -.0in
\end{figure}
In Fig.\ 1 we show the dependence of $\Zv$ on $m_{f}$
(open circle).  The slight quadratic dependence appears because of 
the fact that $V_{\mu}^{\rm conserved}=
\Zv V_{\mu}^{\rm local} + {\cal O}(m_{f}^2a^2)$. 
The quadratic chiral extrapolation gives $\Zv = 0.784(15)$
at $m_{f}=0$, which agrees well with $\Za=0.7777(4)$ \cite{yaoki}. 
The axial-vector renormalization can be obtained from a completely
different calculation of meson two-point correlation functions
based on the relation $\langle A_{\mu}^{\rm conserved}(t){\bar q}
\gamma_{5}q(0)\rangle=\Za\langle A_{\mu}^{\rm local}(t){\bar q}
\gamma_{5}q(0)\rangle$ \cite{rbc00}.

%
%
\begin{figure}[t]
\includegraphics[width=70mm]{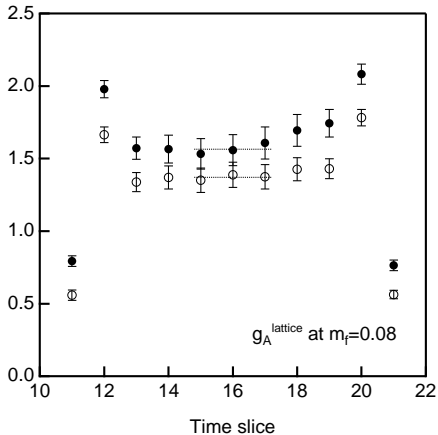}
\vskip -.2in
\caption{The lattice axial charge, $\Ga^{\rm lattice}$,
at $m_{f}=0.08$.  The upper filled and the lower open circles
represent results obtained on larger and
smaller volumes, respectively.  A finite volume
effect is evident between the observed plateaus.
}
\vskip -.2in
\end{figure}
In Fig.\ 2 we plot the axial charge, $\Ga^{\rm lattice}$, against
the location of current insertions at $m_{f}=0.08$. 
Filled and open circle symbols correspond to results from the larger
volume and the smaller volume respectively.  While plateaus 
are evident for $15\leq t \leq 17$ in each case,
$\Ga^{\rm lattice}$ evaluated from the smaller volume is obviously
diminished. 

Next, we evaluate the continuum value of $\Ga$ from
the charge ratios $(\Ga/\Gv)^{\rm lattice}$ averaged 
in the above mentioned time slice range at each $m_{f}$.
To compare to our previous DWF results, we plot 
the continuum value of $\Ga$
versus the square of the $\pi$-$\rho$ mass ratio in Fig.\ 3.
%
%
\begin{figure}[t]
\includegraphics[width=70mm]{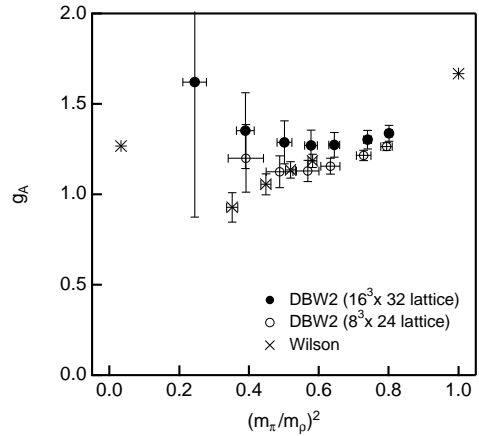}
\vskip -.2in
\caption{$\Ga$ versus $(m_{\pi}/m_{\rho})^2$.
Preliminary results ($\bullet$ and $\circ$) from DBW2 gauge action 
($\beta =0.87$) and
two different physical volumes shows the existence of
a significant finite volume effect.  Our previous results
($\times$) using the Wilson gauge action ($\beta =6.0$) and a physical volume
which is roughly the same as the smaller DBW2 lattice also appears 
to be affected by finite volume. 
}
\vskip -.2in
\end{figure}
Our new results using DBW2 gauge action 
are represented by filled and open circles.
The upper points ($\bullet$) and the lower points
($\circ$) are calculated in the larger spatial volume 
$\sim (2.4 {\rm fm})^3$ and the smaller spatial volume
$\sim (1.2 {\rm fm})^3$ respectively.
There is a clear finite volume effect, which seems to become
large in the lighter quark mass region.  In addition, 
the larger volume results have a mild quark mass dependence 
except for the two lightest points, which still have large 
statistical errors.  On the other hand, 
the smaller volume results seem to be in
rough agreement with our previous results 
using the Wilson gauge action ($\times$) which show
a strong mass dependence. 
As the statistics of our calculation improve, it
will be interesting to compare these results to the recent
continuum calculation by Jaffe~\cite{jaffe}.

In conclusion, using DWF and the DBW2 gauge action,
we have studied the effects of finite physical
volume on $g_A$ by employing a coarse lattice
($a\approx 0.15$ fm) and two lattice sizes ($V\sim (2.4 {\rm 
fm})^3$ and $(1.2 {\rm fm})^3$).
Relevant three-point functions are well behaved (vector, axial
and also tensor).  We confirmed that $\Zv=\Za$
is well satisfied even on this coarse lattice.  We determined the 
continuum value of $\Ga$ in a fully non-perturbative way.  Although we
need more statistics to make a definite conclusion, our preliminary
results suggest that the finite
volume effect is significant. 

We thank RIKEN, Brookhaven National Laboratory and the U.S.\ Department
of Energy for providing the facilities essential for the completion of
this work.

\end{document}